# First-principles study of the structural, electronic, magnetic and ferroelectric properties of a charge ordered Iron(II)-Iron(III) formate framework

Gang Tang,[a] Wei Ren,[c] Jiawang Hong,[a,]*and Alessandro Stroppa[b,]*

[a] School of Aerospace Engineering, Beijing Institute of Technology, Beijing, 100081, China

[b] CNR-SPIN, c/o Dip.to di Scienze Fisiche e Chimiche -Università degli Studi dell'Aquila - Via Vetoio - 67100 - Coppito (AQ), Italy

[c] Department of Physics and International Centre for Quantum and Molecular Structures, Shanghai University, 99 Shangda Road, Shanghai, 200444, China

**Corresponding Author**

*E-mail: hongjw@bit.edu.cn; alessandro.stroppa@spin.cnr.it

**Abstract**

Density functional theory calculations have been performed for the structural, electronic, magnetic and ferroelectric properties of a mixed-valence Fe(II)-Fe(III) formate framework [NH$_2$(CH$_3$)$_2$][Fe$^{III}$Fe$^{II}$(HCOO)$_6$] (DMAFeFe). Recent experiments report a spontaneous electric polarization and our calculations are in agreement with the reported experimental value. Furthermore, we shed light into the microscopic mechanism leading to the observed value and as well how to possibly enhanced the polarization. The interplay between charge ordering, dipolar ordering of DMA$^+$ cations and the induced structural distortions suggest new interesting directions to explore in these complex multifunctional hybrid perovskites.

# 1. Introduction

Metal-organic frameworks (MOFs), consisting of organic cations, a network of metal ions coordinated by organic bridging ligands, have attracted a lot of attention because of their multifunctional properties, such as ferroelectricity, magnetism, and conductivity.[1-17] In particular, the combination of ferroelectric and magnetic orders (*i.e.*, multiferroicity) in a single-phase material is of great technological and fundamental importance.[1-3, 13, 18-21] Furthermore, by considering their diversity in chemical compositions and structural topologies, MOFs are considered as highly promising candidates for multifunctional materials for their potential application in gas absorption and separation, ionic exchange and identification, conductive and catalytic, magnetic and even ferroelectric properties.[4, 22] In particular, some MOFs with perovskite topology showing the coexistence of ferroelectric and magnetic orders, such as $AM^{II}(HCOO)_3$ (A = $NH_2(CH_3)_2$ (DMA) or $NH_4$; M = Mg, Zn, Mn, Ni, Co, or Fe),[5-7, 18-21, 23-26] have been theoretically predicted and/or experimentally characterized. In these studies, it has been shown that the magnetic order originates from metal ions and the ferroelectric order from the organic molecules through a quite peculiar hybrid improper inversion symmetry breaking mechanism. This microscopic interpretation suggests two hybridizing non-polar distortions breaking the inversion symmetry and giving rise to ferroelectric polarization. While the hybrid improper ferroelectricity was initially discussed in the context of inorganic multiferroics,[27] it has become clear that in hybrid materials it may have much more interesting applications and possibility for increasing the functionality of hybrid perovskite.[28] At present, there is an active increasing search for new multifunctional hybrid materials showing large polarization as well as magnetoelectric coupling.

Heterometallic MOFs with a general formula $[NH_2(CH_3)_2][Fe^{III}M^{II}(HCOO)_6]$ (M = Fe, Co, Mn, Mg, Ni, Zn, or Cu), which crystallize in a niccolite-like topology, have been widely investigated.[22, 29-38] The combination of two magnetic metals in the same structure brings rich magnetic properties. In particular, $[NH_2(CH_3)_2][Fe^{III}Fe^{II}(HCOO)_6]$ (DMAFeFe) is the first example exhibiting negative magnetization assigned as Neel N-Type ferrimagnet in a 3D molecule-based magnet.[7] Recently, the magnetoelectric coupling effect in the mixed-valence formate framework was experimentally observed at 3.5 K.[35] According to previous studies,[36] DMAFeFe crystallizes in the trigonal *P*-3$1c$ space group at room temperature. Until 2012, Rodríguez-Carvajal *et al.* first observed an unprecedented order-disorder phase transition in the mixed-valence complex by a neutron diffraction experiment.[37] It was found that DMAFeFe transforms into an antiferroelectric *R*-3*c* phase (see Figure 1a) at low temperature (155 K), involving ordering of $DMA^+$ cations. The reported two phases belong to the centrosymmetric space group, therefore, no ferroelectricity is expected, at least in this temperature range.

However, very recently, Zheng *et al.* found the occurrence of the ferroelectric phase transition from dielectric measurement of DMAFeFe over the low-temperature range (0~100 K).[35] They have measured a ferroelectric polarization of about 0.56 ~ 0.72 $nC/cm^2$ at 2 K, but the crystal structure was not been completely refined yet. However, as the temperature decreases, the DMA$^+$ cations should be ordered in such a way to reduce the symmetry of DMAFeFe structure to a polar space group. Since the low temperature phase has not been fully determined in terms of atomic positions, its structural, electronic, ferroelectric and magnetic properties require further investigations.

In this work, we start from the proposed centrosymmetric structure at T = 93 K and we performed accurate atomic relaxations by using first-principles calculations. We found that, a non-centrosymmetric structure (*R3c* space group) with antiferromagnetic spin ordering as a lower energy compared with the centrosymmetric structure (*R-3c* space group). This proposed low temperature structure shows a ferroelectric polarization as high as 7.41 *n*C/cm$^2$, in reasonable agreement with a recent measurement.[35] This suggests that the DMAFeFe should be a hybrid organic-inorganic multiferroic material at low temperature. We found the small polarization originates mainly from the Fe off-centering of [Fe$^{III}$O$_6$] octahedral. The detailed analysis in terms of the different contributions to electric polarization, due to the interplay between charge ordering, dipolar ordering of DMA$^+$ cations as well as induced structural distortions is the subject of a forthcoming article. Our study suggests possible future directions for engineering multifunctional properties in hybrid organic-inorganic perovskites.

## 2. Computational details

Density-functional theory (DFT) calculations were performed by using the Perdew-Burke-Ernzerhof generalized gradient approximation (PBE-GGA)[39], as implemented in the Vienna Ab initio Simulation Package (VASP)[40-41]. Because of the strong electronic correlation between partially filled $d$ shells, GGA+$U$ method was employed in our calculations[42]. The correlation energy ($U$) and exchange energy ($J$) were chosen to be 4 eV (5 eV) and 0.89 eV (0.89 eV) for $Fe^{2+}$ ($Fe^{3+}$), respectively from previous literature.[43] The plane-wave cut-off energy was set to 500 eV and a 4×4×4 Monkhorst-Pack grid of $k$-points was employed for sampling the Brillouin zone.[44] The atomic positions were relaxed until the force on each atom was smaller than 0.02 eV/Å. For estimating the polarization, the Berry phase theory as implemented in the VASP code was used.[45-46]

## 3. Results and discussion

Firstly, we performed density functional theory simulations to study a possible ground state of DMAFeFe at low temperature. We started with an ordered arrangements of $DMA^+$ cations compatible with the centrosymmetric $R$-$3c$ space group. After relaxing all the atomic positions, we found that the polar structure ($R3c$ phase) has a lower total energy than the non-polar structure ($R$-$3c$ phase) by 73.36 meV/f.u. The atomic coordinates of [FeO$_6$] octahedra in $R$-$3c$ and $R3c$ phase are summarized in Table 1. The polarization is comparable with the experimental result (see below). Figure 1 shows the structures representing the centrosymmetric reference phase ($R$-$3c$) and the proposed ferroelectric phase ($R3c$) for DMAFeFe. In both cases, the metal centers ($Fe^{II}$ and $Fe^{III}$) are bridged by formate ions in the *anti-anti* mode configuration, and $DMA^+$ organic ions are filled in the cavities of the 3D framework. Both metal centers are surrounded by six oxygen atoms of formate bridging ligands to form an octahedral coordination.

Starting from the experimental lattice parameters ($a = b = 14.26$ Å, $c = 41.44$ Å; $\alpha = \beta = 90°$, $\gamma = 120°$),[37] only the atomic positions of two structures were relaxed. And the corresponding bond lengths were summarized in Table 2.

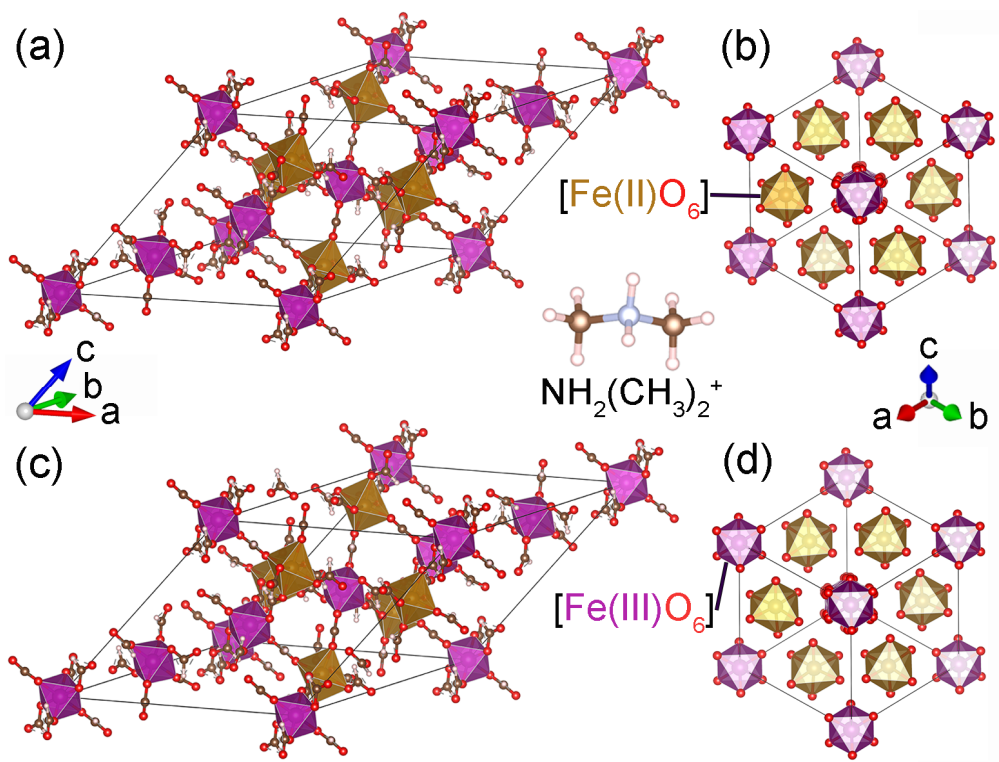

**Figure 1.** Schematic view of the primitive cell of (a, b) *R-3c* phase, (c, d) *R3c* phase [NH$_2$(CH$_3$)$_2$][Fe$^{III}$Fe$^{II}$(HCOO)$_6$]. For clarity, the dimethylammonium ions (DMA$^+$) are omitted in the structure.

**Table 1.** Atom coordinates of [FeO$_6$] octahedra in relaxed *R-3c* and *R3c* phase of [NH$_2$(CH$_3$)$_2$][Fe$^{III}$Fe$^{II}$(HCOO)$_6$].

|  | Atom | Wyckoff Position | Site Symmetry | x | y | z |
| --- | --- | --- | --- | --- | --- | --- |
|  | Fe1 | 18e | 2 | 0.67085 | 0 | -0.25 |
|  | Fe2 | 6b | -3 | 1/3 | -1/3 | -1/3 |
|  | Fe3 | 12c | 3 | 1/3 | -1/3 | 0.00206 |
| *R-3c* | O1 | 36f | 1 | 0.46570 | -0.25183 | -0.02696 |
|  | O2 | 36f | 1 | 0.41143 | -0.39073 | 0.02835 |
|  | O3 | 36f | 1 | 0.74943 | 0.14021 | -0.27964 |
|  | O4 | 36f | 1 | 0.74114 | -0.08223 | -0.27458 |
| | O5 | 36f | 1 | 0.52802 | -0.07250 | -0.27811 |
| | O6 | 36f | 1 | 0.39398 | -0.20176 | -0.30694 |
| | Fe1 | 6a | 3 | 0 | 0 | -0.50023 |
| | Fe2 | 6a | 3 | 0 | 0 | -0.16413 |
| | Fe3 | 6a | 3 | 0 | 0 | -0.33596 |
| | Fe4 | 18b | 1 | 0.34138 | 1/3 | -0.41667 |
| | O1 | 18b | 1 | 0.13263 | 0.08232 | -0.19313 |
| | O2 | 18b | 1 | 0.08266 | 0.13256 | -0.30693 |
| | O3 | 18b | 1 | 0.08063 | -0.05449 | -0.13761 |
| R3c | O4 | 18b | 1 | -0.05476 | 0.08045 | -0.36249 |
| | O5 | 18b | 1 | 0.42391 | 0.47983 | -0.44519 |
| | O6 | 18b | 1 | -0.18775 | 0.08898 | -0.38829 |
| | O7 | 18b | 1 | 0.40603 | 0.24896 | -0.44238 |
| | O8 | 18b | 1 | -0.08333 | -0.26113 | -0.72402 |
| | O9 | 18b | 1 | 0.20560 | 0.26829 | -0.44641 |
| | O10 | 18b | 1 | -0.39903 | -0.13061 | -0.38816 |
| | O11 | 18b | 1 | 0.06669 | 0.13541 | -0.47274 |
| | O12 | 18b | 1 | -0.53132 | -0.26776 | -0.36105 |

**Table 2.** Calculated bond lengths and local magnetic moments in [FeO$_6$] octahedra for R-3c and R3c phase of [NH$_2$(CH$_3$)$_2$][Fe$^{III}$Fe$^{II}$(HCOO)$_6$].

| | Bond length (Å) | | Magnetic moment ($\mu_B$) | | |
|---|---|---|---|---|---|
| | R-3c (167) | R3c (161) | | R-3c (167) | R3c (161) |
| Fe(II)-O | 2.097-2.150 | 2.076-2.155 | Fe(II) | 3.7 | 3.7 |
| Fe(III)-O (1) | 2.019 | 2.019-2.020 | | | |
| Fe(III)-O (2) | 2.013 | 2.005-2.039 | Fe(III) | -4.3 | -4.3 |
| Fe(III)-O (3) | 2.028 | 2.003-2.039 | | | |

From Table 2, it can be noted that both structures have one kind of [Fe$^{II}$O$_6$] octahedra and three kinds of [Fe$^{III}$O$_6$] octahedra. [Fe$^{II}$O$_6$] octahedra are distorted in both phases, as reflected by the distinct Fe(II)-O bond lengths along the different directions. However, [Fe$^{III}$O$_6$] octahedra are only distorted in the R3c phase, consistent with the reduced symmetry. In addition, the Fe-O bond lengths in the two phases are very close. Moreover, the bond lengths of Fe(II)-O are longer than that of Fe(III)-O. This assignment is further supported by the larger ionic size of $Fe^{2+}$ (0.78 Å) relative to $Fe^{3+}$ (0.645 Å).[47]

We also evaluate the ferroelectric polarization by using the Berry phase theory.[45-46] Starting from the low-symmetry 3.5 K structure ($\lambda = 1$) with space group $R3c$, we consider the structure with space group $R\text{-}3c$ as the reference structure ($\lambda = 0$) and we build a one-to-one atomic mapping between the $R3c$ and $R\text{-}3c$ structures. The atomic distortion field connecting the two structures have been parametrized by $\lambda$ representing the normalized amplitude of the distortion field. Figure 2 shows the evolution of total energy difference and the polarization from the $R3c$ to $R\text{-}3c$ structure, as a function of the normalized amplitude ($\lambda$) of atomic displacements. The estimated total polarization along the (111) direction (see Figure 1) is 7.41 $nC/cm^2$, which is larger than the experimentally measured polarization values (0.56~0.72 $nC/cm^2$). The discrepancy may be due to the fact that the measured values are related to temperature and magnetic field used in the experiments.[35]

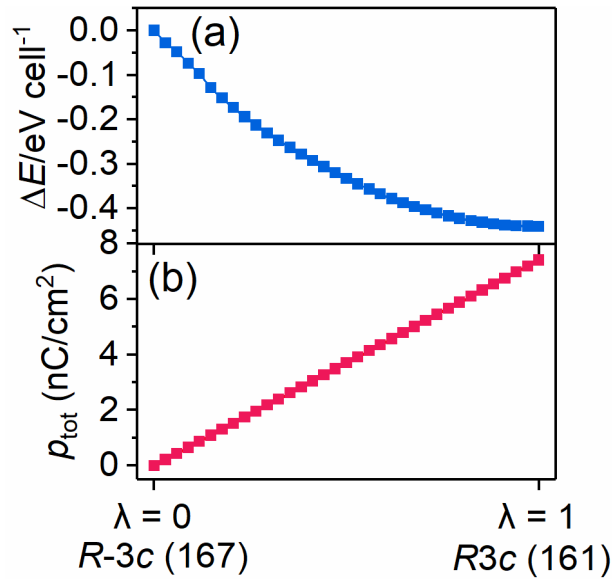

**Figure 2.** Path connecting the centrosymmetric structure ($\lambda = 0$) and the polar structure ($\lambda = 1$). (a) Total energy difference with respect to the paraelectric phase. (b) Total polarization along the (111) directions as a function $\lambda$.

Next, we investigated the magnetic properties of low-temperature phase DMAFeFe. As revealed by previous experiment,[37] the magnetic structure below 37 K corresponds to a weakly noncollinear ferrimagnetic structure, where the spin moment of the $Fe^{II}$ and $Fe^{III}$ sites are antiferromagnetically coupled along the *c*-axis (see Figure 3a). Since the structure is very complex, consisting of 222 atoms per unit cell, in our study we considered only collinear spins. On the basis of antiferromagnetic configuration, the calculated local magnetic moments were shown in Table 2. For *R*-3*c* and *R*3*c* structures, the spin moments on $Fe^{II}$ and $Fe^{III}$ atoms are 3.7 $\mu_B$ and -4.7 $\mu_B$, respectively, in agreement with the experimentally measured values (3.9/-4.1 $\mu_B$).[37] Figure 3b shows the spin charge density distributions for *R*3*c* structure with localized spin-density at $Fe^{II}$ and $Fe^{III}$ sites with opposite spin directions, further supporting our magnetic calculation. The local magnetic moment on $Fe^{II}$ and $Fe^{III}$ atoms can be well understood from their corresponding *d* electron configurations (see Figure 3c). $Fe^{2+}$ ($Fe^{3+}$) has $d^6$ ($d^5$) configuration and there are four (five) unpaired *d* electrons, providing the largest local magnetic moment of 3.7 $\mu_B$ (4.7 $\mu_B$).

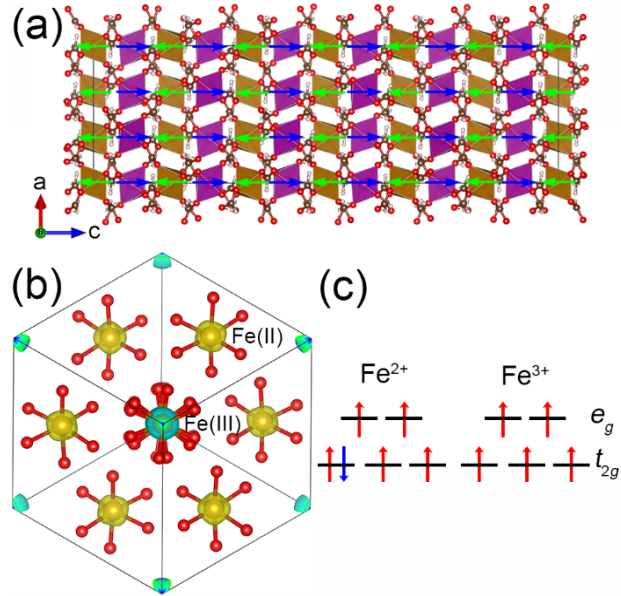

**Figure 3.** (a) View along the *b*-axis of the unit cell of $[NH_2(CH_3)_2][Fe^{III}Fe^{II}(HCOO)_6]$ together with the arrangement of the spins of each $Fe^{II}$ and $Fe^{III}$ site, orange/green and purple/yellow, respectively. (b) spin charge density for *R3c* phase $[NH_2(CH_3)_2][Fe^{III}Fe^{II}(HCOO)_6]$ and (c) *d* electron configurations of $Fe^{2+}$ and $Fe^{3+}$. The isosurface value is set to 0.448 e Å$^{-3}$. Note that the yellow and green parts correspond to electron density increase and decrease, respectively.

Finally, we examined the electronic properties of *R3c* structure. As shown in Figure 4 and 5, PBE+*U* method predicts an indirect band gap of 0.74 eV, the valence band maximum mainly consists of spin down-based orbitals of Fe(II) 3d and O 2p, and the conduction band minimum is mainly of spin up-based orbitals of Fe(III) 3d and O 2p. To our knowledge, at present, no solid-state conductivity and optical band gap measurements of DMAFeFe have been reported. Although containing mixed-valence Fe(II)-Fe(III) cations, due to the large crystallographic unit cell and relatively isolated [FeO$_6$] octahedra, no apparent band dispersion can be observed, indicating localized carriers and low levels of conductivity in *R3c* phase DMAFeFe. Meanwhile, it is worth noting that the observed band characters are very similar to MOF-5 (Zn4O(1,4-dicarboxylate)3).[48]

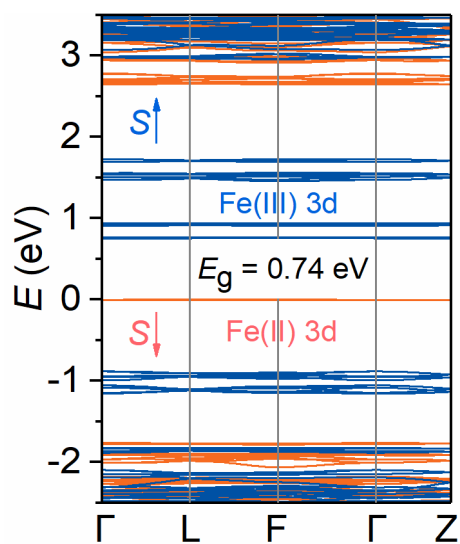

**Figure 4.** Band structure for $R3c$ phase [NH$_2$(CH$_3$)$_2$][Fe$^{III}$Fe$^{II}$(HCOO)$_6$].

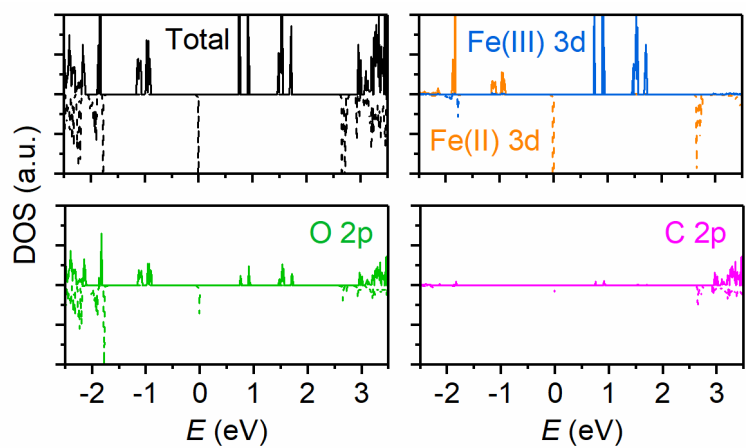

**Figure 5.** Total and partial density of states for $R3c$ phase [NH$_2$(CH$_3$)$_2$][Fe$^{III}$Fe$^{II}$(HCOO)$_6$]

## 4. Conclusions

In summary, we reported the structural, electronic, magnetic and ferroelectric properties of a mixed-valence Fe(II)-Fe(III) formate framework by using the first-principles calculations. Our results imply a possibly new ferroelectric phase with the *R*3*c* space group below 3.5 K could reasonably explain the low temperature ferroelectricity of DMAFeFe observed in recent experiment. More detailed study is in progress in order to disentangle the different contributions to the electric polarization, but this goes beyond the purpose of the present report. Finally, the results of electronic properties show that no apparent band dispersion can be observed for *R*3*c* phase DMAFeFe, indicating localized carriers and low levels of conductivity in the compound. Our observations provide a possibility to design a multifunctional mixed-valence MOF.

**Acknowledgments**

This work is supported by the National Science Foundation of China (Grant No. 11572040), the Thousand Young Talents Program of China, and Graduate Technological Innovation Project of Beijing Institute of Technology. Theoretical calculations were performed using resources of the National Supercomputer Centre in Guangzhou, which is supported by Special Program for Applied Research on Super Computation of the NSFC-Guangdong Joint Fund (the second phase) under Grant No. U1501501.

**References**

1. Spaldin, N.; Ramesh, R., Advances in Magnetoelectric Multiferroics. *Nat. Mater.* **2019**, *18*, 203.

2. Stroppa, A.; Barone, P.; Jain, P.; Perez‑Mato, J.; Picozzi, S., Hybrid Improper Ferroelectricity in a Multiferroic and Magnetoelectric Metal-Organic Framework. *Adv. Mater.* **2013**, *25*, 2284-2290.

3. Stroppa, A.; Jain, P.; Barone, P.; Marsman, M.; Perez‑Mato, J. M.; Cheetham, A. K.;

Kroto, H. W.; Picozzi, S., Electric Control of Magnetization and Interplay between Orbital Ordering and Ferroelectricity in a Multiferroic Metal-Organic Framework. *Angew. Chem. Int. Ed.* **2011**, *50*, 5847-5850.

4. Yuan, S.; Feng, L.; Wang, K.; Pang, J.; Bosch, M.; Lollar, C.; Sun, Y.; Qin, J.; Yang, X.; Zhang, P., Stable Metal-Organic Frameworks: Design, Synthesis, and Applications. *Adv. Mater.* **2018**, *30*, 1704303.

5. Xu, G.-C.; Zhang, W.; Ma, X.-M.; Chen, Y.-H.; Zhang, L.; Cai, H.-L.; Wang, Z.-M.; Xiong, R.-G.; Gao, S., Coexistence of Magnetic and Electric Orderings in the Metal-Formate Frameworks of [NH$_4$][M(HCOO)$_3$]. *J. Am. Chem. Soc.* **2011**, *133*, 14948-14951.

6. Mączka, M.; Gagor, A.; Macalik, B.; Pikul, A.; Ptak, M.; Hanuza, J., Order-Disorder Transition and Weak Ferromagnetism in the Perovskite Metal Formate Frameworks of [(CH$_3$)$_2$NH$_2$][M(HCOO)$_3$] and [(CH3)$_2$ND$_2$][M(HCOO)$_3$](M = Ni, Mn). *Inorg. Chem.* **2013**, *53*, 457-467.

7. Jain, P.; Ramachandran, V.; Clark, R. J.; Zhou, H. D.; Toby, B. H.; Dalal, N. S.; Kroto, H. W.; Cheetham, A. K., Multiferroic Behavior Associated with an Order-Disorder Hydrogen Bonding Transition in Metal-Organic Frameworks (MOFs) with the Perovskite ABX$_3$ Architecture. *J. Am. Chem. Soc.* **2009**, *131*, 13625-13627.

8. Boström, H. L.; Collings, I. E.; Cairns, A. B.; Romao, C. P.; Goodwin, A. L., High-Pressure Behaviour of Prussian Blue Analogues: Interplay of Hydration, Jahn-Teller Distortions and Vacancies. *Dalton Trans.* **2019**, *48*, 1647-1655.

9. Ji, L.-J.; Sun, S.-J.; Qin, Y.; Li, K.; Li, W., Mechanical Properties of Hybrid Organic-Inorganic Perovskites. *Coord. Chem. Rev.* **2019**, *391*, 15-29.

10. Mączka, M.; Ptak, M.; Gągor, A.; Sieradzki, A.; Peksa, P.; Usevicius, G.; Simenas, M.; Leite, F. F.; Paraguassu, W., Temperature-and Pressure-Dependent Studies of a Highly Flexible and Compressible Perovskite-Like Cadmium Dicyanamide Framework Templated with Protonated Tetrapropylamine. *J. Mater. Chem. C* **2019**, *7*, 2408-2420.

11. Phillips, A. E., Rigid Units Revisited. *Acta Crystallogr., Sect. A: Found. Adv.* **2018**, *74*, 406-407.

12. Phillips, A. E., Introduction: Minerals to Metal-Organic Frameworks. *Philos. Trans. R. Soc., A* http://doi.org/10.1098/rsta.2019.0153.

13. Tian, W.; Cao, H.; Clune, A. J.; Hughey, K. D.; Hong, T.; Yan, J.-Q.; Agrawal, H. K.; Singleton, J.; Sales, B. C.; Fishman, R. S., Electronic Phase Separation and Magnetic-Field-Induced Phenomena in Molecular Multiferroic $(ND_4)_2FeCl_5D_2O$. *Phys. Rev. B* **2018**, *98*, 054407.

14. Wei, W.; Li, W.; Butler, K. T.; Feng, G.; Howard, C. J.; Carpenter, M. A.; Lu, P.; Walsh, A.; Cheetham, A. K., An Unusual Phase Transition Driven by Vibrational Entropy Changes in a Hybrid Organic-Inorganic Perovskite. *Angew. Chem. Int. Ed.* **2018**, *130*, 9070-9074.

15. Xiong, Y.-A.; Sha, T.-T.; Pan, Q.; Song, X.-J.; Miao, S.-R.; Jing, Z.-Y.; Feng, Z.-J.; You, Y.-M.; Xiong, R.-G., The First Nickel (II)-Nitrite-Based Molecular Perovskite Ferroelectric. *Angew. Chem. Int. Ed.* **2019**.

16. Yang, Z.; Cai, G.; Bull, C. L.; Tucker, M. G.; Dove, M. T.; Friedrich, A.; Phillips, A. E., Hydrogen-Bond-Mediated Structural Variation of Metal Guanidinium Formate Hybrid Perovskites under Pressure. *Philos. Trans. R. Soc.* **2019**.

17. You, Y.-M.; Liao, W.-Q.; Zhao, D.; Ye, H.-Y.; Zhang, Y.; Zhou, Q.; Niu, X.; Wang, J.; Li, P.-F.; Fu, D.-W., An Organic-Inorganic Perovskite Ferroelectric with Large Piezoelectric Response. *Science* **2017**, *357*, 306-309.

18. Abhyankar, N.; Kweon, J. J.; Orio, M.; Bertaina, S.; Lee, M.; Choi, E. S.; Fu, R.; Dalal, N. S., Understanding Ferroelectricity in the Pb-Free Perovskite-Like Metal–Organic Framework $[(CH_3)_2NH_2]Zn(HCOO)_3$: Dielectric, 2D NMR, and Theoretical Studies. *J. Phys. Chem. C* **2017**, *121*, 6314-6322.

19. Hughey, K. D.; Clune, A. J.; Yokosuk, M. O.; al-Wahish, A.; O'Neal, K. R.; Fan, S.; Abhyankar, N.; Xiang, H.; Li, Z.; Singleton, J., Phonon Mode Links Ferroicities in Multiferroic $[(CH_3)_2NH_2]Mn(HCOO)_3$. *Phys. Rev. B* **2017**, *96*, 180305.

20. Hughey, K. D.; Clune, A. J.; Yokosuk, M. O.; Li, J.; Abhyankar, N.; Ding, X.; Dalal, N. S.; Xiang, H.; Smirnov, D.; Singleton, J., Structure-Property Relations in Multiferroic $[(CH_3)_2NH_2]M(HCOO)_3$ (M = Mn, Co, Ni). *Inorg. Chem.* **2018**, *57*, 11569-

11577.

21. Mazzuca, L.; Cañadillas-Delgado, L.; Fabelo, O.; Rodríguez-Velamazán, J. A.; Luzon, J.; Vallcorba, O.; Simonet, V.; Colin, C. V.; Rodríguez-Carvajal, J., Microscopic Insights on the Multiferroic Perovskite-Like [CH$_3$NH$_3$][Co(COOH)$_3$] Compound. *Chem. - Eur. J.* **2018**, *24*, 388-399.

22. Abednatanzi, S.; Derakhshandeh, P. G.; Depauw, H.; Coudert, F.-X.; Vrielinck, H.; Van Der Voort, P.; Leus, K., Mixed-Metal Metal-Organic Frameworks. *Chem. Soc. Rev.* **2019**, *48*, 2535-2565.

23. Clune, A.; Hughey, K.; Lee, C.; Abhyankar, N.; Ding, X.; Dalal, N.; Whangbo, M.-H.; Singleton, J.; Musfeldt, J., Magnetic Field-Temperature Phase Diagram of Multiferroic [(CH$_3$)$_2$NH$_2$]Mn(HCOO)$_3$. *Phys. Rev. B* **2017**, *96*, 104424.

24. Nakayama, Y.; Nishihara, S.; Inoue, K.; Suzuki, T.; Kurmoo, M., Coupling of Magnetic and Elastic Domains in the Organic-Inorganic Layered Perovskite-Like (C$_6$H$_5$C$_2$H$_4$NH$_3$)$_2$Fe$^{II}$Cl$_4$ Crystal. *Angew. Chem. Int. Ed.* **2017**, *129*, 9495-9498.

25. Ptak, M.; Dziuk, B.; Stefańska, D.; Hermanowicz, K., The Structural, Phonon and Optical Properties of [CH$_3$NH$_3$]M$_{0.5}$Cr$_x$Al$_{0.5-x}$(HCOO)$_3$ (M = Na, K; $x$ = 0, 0.025, 0.5) Metal-Organic Framework Perovskites for Luminescence Thermometry. *Phys. Chem. Chem. Phys.* **2019**.

26. Wang, Z.; Jain, P.; Choi, K.-Y.; van Tol, J.; Cheetham, A. K.; Kroto, H. W.; Koo, H.-J.; Zhou, H.; Hwang, J.; Choi, E. S., Dimethylammonium Copper Formate [(CH$_3$)$_2$NH$_2$]Cu(HCOO)$_3$: A Metal-Organic Framework with Quasi-One-Dimensional Antiferromagnetism and Magnetostriction. *Phys. Rev. B* **2013**, *87*, 224406.

27. Benedek, N. A.; Fennie, C. J., Hybrid Improper Ferroelectricity: A Mechanism for Controllable Polarization-Magnetization Coupling. *Phys. Rev. Lett.* **2011**, *106*, 107204.

28. Boström, H. L.; Senn, M. S.; Goodwin, A. L., Recipes for Improper Ferroelectricity in Molecular Perovskites. *Nat. Commun.* **2018**, *9*, 2380.

29. Ciupa, A.; Mączka, M.; Gągor, A.; Sieradzki, A.; Trzmiel, J.; Pikul, A.; Ptak, M., Temperature-Dependent Studies of [(CH$_3$)$_2$NH$_2$][Fe$^{III}$M$^{II}$(HCOO)$_6$] Frameworks (M$^{II}$ = Fe and Mg): Structural, Magnetic, Dielectric and Phonon Properties. *Dalton Trans.*

**2015**, *44*, 8846-8854.

30. Guo, J. B.; Chen, L. H.; Ke, H.; Wang, X.; Zhao, H. X.; Long, L. S.; Zheng, L. S., Dielectric Tunability, Expanding the Function of Metal-Organic Frameworks. *Phys. Status Solidi RRL* **2018**, *12*, 1700425.

31. Mazzuca, L.; Cañadillas-Delgado, L.; Rodríguez-Velamazán, J. A.; Fabelo, O.; Scarrozza, M.; Stroppa, A.; Picozzi, S.; Zhao, J.-P.; Bu, X.-H.; Rodríguez-Carvajal, J., Magnetic Structures of Heterometallic M(II)-M(III) Formate Compounds. *Inorg. Chem.* **2016**, *56*, 197-207.

32. Yang, L.; Li, J.; Pu, T.-C.; Kong, M.; Zhang, J.; Song, Y., Study of the Relationship between Magnetic Field and Dielectric Properties in Two Ferromagnetic Complexes. *RSC Adv.* **2017**, *7*, 47913-47919.

33. Sieradzki, A.; Pawlus, S.; Tripathy, S.; Gągor, A.; Ciupa, A.; Mączka, M.; Paluch, M., Dielectric Relaxation Behavior in Antiferroelectric Metal Organic Framework [(CH$_3$)$_2$NH$_2$][Fe$^{III}$Fe$^{II}$(HCOO)$_6$] Single Crystals. *Phys. Chem. Chem. Phys.* **2016**, *18*, 8462-8467.

34. Zhao, J.-P.; Hu, B.-W.; Lloret, F.; Tao, J.; Yang, Q.; Zhang, X.-F.; Bu, X.-H., Magnetic Behavior Control in Niccolite Structural Metal Formate Frameworks [NH$_2$(CH$_3$)$_2$][Fe$^{III}$M$^{II}$(HCOO)$_6$] (M = Fe, Mn, and Co) by Varying the Divalent Metal Ions. *Inorg. Chem.* **2010**, *49*, 10390-10399.

35. Guo, J.; Chen, L.; Li, D.; Zhao, H.; Dong, X.; Long, L.; Huang, R.; Zheng, L., An Insight into the Magnetoelectric Coupling Effect in the MOF of [NH$_2$(CH$_3$)$_2$]$_n$[Fe$^{III}$Fe$^{II}$(HCOO)$_6$]$_n$. *Appl. Phys. Lett.* **2017**, *110*, 192902.

36. Hagen, K. S.; Naik, S. G.; Huynh, B. H.; Masello, A.; Christou, G., Intensely Colored Mixed-Valence Iron (II) Iron (III) Formate Analogue of Prussian Blue Exhibits Néel N-Type Ferrimagnetism. *J. Am. Chem. Soc.* **2009**, *131*, 7516-7517.

37. Cañadillas-Delgado, L.; Fabelo, O.; Rodríguez-Velamazán, J. A.; Lemée-Cailleau, M.-H. l. n.; Mason, S. A.; Pardo, E.; Lloret, F.; Zhao, J.-P.; Bu, X.-H.; Simonet, V., The Role of Order–Disorder Transitions in the Quest for Molecular Multiferroics: Structural and Magnetic Neutron Studies of a Mixed Valence Iron(II)-Iron(III) Formate

Framework. *J. Am. Chem. Soc.* **2012**, *134*, 19772-19781.

38. Mączka, M.; Kucharska, E.; Gągor, A.; Pikul, A.; Hanuza, J., Synthesis, Magnetic and Vibrational Properties of Two Novel Mixed-Valence Iron(II)-Iron(III) Formate Frameworks. *J. Solid State Chem.* **2018**, *258*, 163-169.

39. Perdew, J. P.; Burke, K.; Ernzerhof, M., Generalized Gradient Approximation Made Simple. *Phys. Rev. Lett.* **1996**, *77*, 3865.

40. Kresse, G.; Furthmüller, J., Efficient Iterative Schemes for Ab Initio Total-Energy Calculations Using a Plane-Wave Basis Set. *Phys. Rev. B* **1996**, *54*, 11169.

41. Kresse, G.; Joubert, D., From Ultrasoft Pseudopotentials to the Projector Augmented-Wave Method. *Phys. Rev. B* **1999**, *59*, 1758.

42. Anisimov, V. I.; Aryasetiawan, F.; Lichtenstein, A., First-Principles Calculations of the Electronic Structure and Spectra of Strongly Correlated Systems: The LDA+U Method. *J. Phys.: Condens. Matter* **1997**, *9*, 767.

43. Zhang, Q.; Li, B.; Chen, L., First-Principles Study of Microporous Magnets M-MOF-74 (M = Ni, Co, Fe, Mn): the Role of Metal Centers. *Inorg. Chem.* **2013**, *52*, 9356-9362.

44. Monkhorst, H. J.; Pack, J. D., Special Points for Brillouin-Zone Integrations. *Phys. Rev. B* **1976**, *13*, 5188.

45. King-Smith, R.; Vanderbilt, D., Theory of Polarization of Crystalline Solids. *Phys. Rev. B* **1993**, *47*, 1651.

46. Resta, R., Macroscopic Polarization in Crystalline Dielectrics: The Geometric Phase Approach. *Rev. Mod. Phys.* **1994**, *66*, 899.

47. Shannon, R. D., Revised Effective Ionic Radii and Systematic Studies of Interatomic Distances in Halides and Chalcogenides. *Acta Crystallogr., Sect. A: Cryst. Phys., Diffr., Theor. Gen. Crystallogr.* **1976**, *32*, 751-767.

48. Hendon, C. H.; Tiana, D.; Walsh, A., Conductive Metal-Organic Frameworks and Networks: Fact or Fantasy? *Phys. Chem. Chem. Phys.* **2012**, *14*, 13120-13132.